\documentclass[conference,10pt,letterpaper]{IEEEtran}
\usepackage[T1]{fontenc}
\usepackage[utf8]{inputenc}
\usepackage{graphicx,booktabs,array,tabularx,url}
\usepackage[hidelinks]{hyperref}
\usepackage{eso-pic}
\graphicspath{{figures/}}
\hypersetup{pdftitle={Operating an Agent-Driven Dataset Factory: Discovery, Construction, and Public Release},pdfauthor={Eduardo Arino de la Rubia and Szilard Pafka}}
\title{Operating an Agent-Driven Dataset Factory:\\Discovery, Construction, and Public Release}
\author{\IEEEauthorblockN{Eduardo Ari\~no de la Rubia}\IEEEauthorblockA{Central European University\\\texttt{rubiae@ceu.edu}}\and\IEEEauthorblockN{Szilard Pafka}\IEEEauthorblockA{Epoch\\\texttt{szilard@epoch.com}}}
\begin{document}
\AddToShipoutPictureFG*{\AtPageLowerLeft{\put(\LenToUnit{0.5in},\LenToUnit{0.22in}){\parbox{\dimexpr\paperwidth-1in\relax}{\footnotesize\centering This work has been submitted to the IEEE for possible publication. Copyright may be transferred without notice, after which this version may no longer be accessible.}}}}
\bstctlcite{IEEEexample:BSTcontrol}
\maketitle

\begin{abstract}
Turning public data into prediction datasets requires connecting source discovery, task definition, construction, and distribution. We deployed Dataset Factory to perform these steps continuously under a standing release policy. Bounded language-model tasks interpret sources, write construction code, and review task meaning; deterministic software manages state, workers, retries, and publication. Over 81.53 hours, the factory published 14 datasets at an estimated recurring operating expenditure of about \$2.74 per publication. The accounting includes failed attempts, inference, workers, and hosting. Separate agents completed prediction experiments on all twelve evaluated releases. Source review nevertheless identified a task whose construction filters contradicted its population declaration. The original reviewer accepted later-record exclusions without reconciling them with the published selection rule. This experience documents an operating source-to-release workflow and shows why successful execution and consumer use need separate checks of task meaning.
\end{abstract}
\begin{IEEEkeywords}agentic engineering, dataset construction, deployment experience, operating cost, semantic review\end{IEEEkeywords}

\section{Introduction}
Dataset Factory connected public-source discovery to fourteen published prediction datasets in an 81.53-hour observation window. Separate agents used twelve for modeling experiments. Automated acceptance also passed a task whose construction contradicted its population declaration.

We built the factory for a small research group that needs unfamiliar, real-world prediction tasks. Public records require interpretation, filtering, labels, splits, and reproducible packaging before another researcher can use them. Analyst interviews and curation records document this work in discovery, transformation, and quality checks~\cite{kandel,lafia}. Automating a single step still leaves the problem of connecting the steps across sources.

This report describes the model/controller division of work, production and consumer use, and lessons from observed failures. Its evidence includes repeated attempts, recurring expenditure, and the original inputs and response of the failed semantic review.

\section{System and Work Allocation}
\subsection{Discovery and construction}
Each release supplies a binary prediction task, training/validation/holdout splits, a predictor contract, construction code, a baseline, and source and license evidence. Matching versioned artifacts are published on GitHub and Hugging Face. The factory's code is public.\footnote{\url{https://github.com/earino/dataset-factory}}

Models interpret sources and generate code. The controller owns persistent state, resource limits, and external operations, so failed attempts can resume from recorded state.

\begin{figure*}[t]
\centering\includegraphics[width=\textwidth]{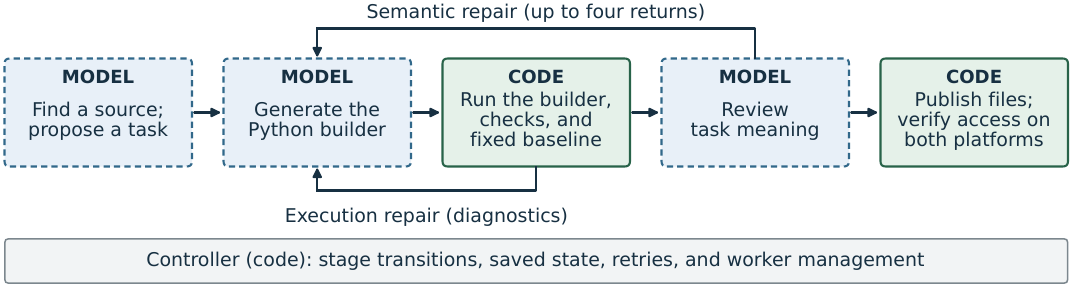}
\caption{Source-to-release workflow. Blue dashed boxes are model tasks; green solid boxes are code operations. Execution and semantic repairs return to builder generation.}
\label{fig:workflow}
\end{figure*}

A scheduled controller advances SQLite state through the four stages in Fig.~\ref{fig:workflow}, keeping up to two candidate slots pending. Each receives previous task titles and a domain focus alternating infrastructure/environment with institutions, commerce, transport, and public services.

The discovery model uses web search and extraction to inspect downloads, field definitions, and redistribution terms. It proposes source evidence, target, population, prediction time, available features, and chronological splits, or defers with a reason. Code checks required fields and duplicate task identifiers before advancing the model-selected source and task.

Construction is a separate model task without tools. It receives the source proposal and a descriptor schema, then returns Python code and task metadata. The builder writes the three splits and source checks. A temporary worker executes it, checks the full artifacts, and runs a fixed baseline. Checks cover file/schema compatibility, labels, splits, task-specific leakage checks, source evidence, and reproducibility records. Current-vintage retrospective tasks are permitted with declared assumptions; historical forecasting requires evidence of past input availability.

\subsection{Review and bounded repair}
The semantic reviewer receives the source/task description, complete builder, construction notes, evaluation contract, and compact source, license, qualification, and baseline reports. It can consult publisher documentation and returns accept, repair, or reject with specific findings. Mechanical checks cannot be waived by that response. The review packet contains selected descriptor fields rather than the complete published descriptor, a distinction that matters in the failure examined in Section~\ref{sec:validity}.

Repairs receive saved proposals and bounded execution diagnostics or semantic findings, reusing unchanged fields. Access or terms failures can invoke same-source web research; malformed responses can be repaired without rediscovery. Rebuilt artifacts repeat checks and review. The controller permits up to four semantic returns to construction before deferral. Model-task and worker failures have separate five-failure limits per stage.

Bounded model sessions run in Hermes Agent\footnote{\url{https://github.com/NousResearch/hermes-agent}}, using DeepSeek V4.1 Flash through Nous and Ollama Cloud. Research and review were capped at twelve tool-calling cycles, each comprising a model response and execution of its requested web tools. All tasks had a ten-minute timeout. Construction returns code for worker execution; it cannot run that code during generation. Models cannot directly launch workers, change coordinator state, or publish releases.

\subsection{Publication and autonomy}
One coordinator launches workers, preserves operation identities, and reconciles results; independent cleanup removes expired workers. A manifest binds accepted files to checksums. Workers retain source hashes and terms evidence; releases preserve upstream rights and attribution. Publication succeeds only when code validates anonymous download receipts from both platforms against the release identity and expected row counts. Saved attempts and frozen releases preserve evidence for resumption and inspection.

A standing policy permits publication after artifact, semantic, and distribution checks, without per-dataset human approval. During the observation window, service administration changed the inference route and repaired worker-budget accounting. The inspected records show no manual changes to source selection, dataset construction, labels, or acceptance decisions.

\begin{table*}[t]
\caption{Publications in verification order: principal review findings and disclosure in pinned release documents.}
\label{tab:releases}
\centering\small
\renewcommand{\arraystretch}{1.17}
\setlength{\tabcolsep}{4pt}
\begin{tabular}{@{}>{\raggedright\arraybackslash}p{.25\textwidth}>{\raggedright\arraybackslash}p{.48\textwidth}>{\raggedright\arraybackslash}p{.235\textwidth}@{}}
\toprule
Task (domain) & Principal review finding & Release disclosure\\\midrule
\csname @@input\endcsname generated/release-rows.tex
\bottomrule
\end{tabular}
\end{table*}

\section{Observation and Evaluation Method}
\subsection{Production and expenditure}
Hosted research began in September 2026; the controller described here replaced earlier workflows in November. Observation began with enrollment of new publications on 12 November at 21:28 UTC and ended on 16 November at 07:00 UTC, spanning 81.53 hours. Seven already-published pilot releases were excluded. A read-only export retains candidates, attempts, usage, and publication identities. A candidate initially denotes a search slot. Milestones count distinct candidates; each attempt is a durable stage execution. Waiting resumes the same attempt. Retries record unsuccessful operations or reconciliation, not unique defects.

Resource-based expenditure estimates use token categories, dated provider tariffs, worker lifetimes, and hosting, without invoice reconciliation. Hetzner CPX32 workers had four allocated CPUs, a 30-minute task timeout, and a 60-minute lifetime limit. Inference uses peak/off-peak timing, with all-peak sensitivity; Nous-routed tokens use direct DeepSeek rates. Workers use whole billed hours; hosting is allocated from a \$20 monthly plan.

Accounting includes unsuccessful production work and whole recorded session/worker lifetimes crossing the opening boundary, with worker exposure capped at close. It excludes development, human labor, upstream data collection, unmetered services, and the separate consumer evaluation. Expenditure divided by publications verified within the window is an operating ratio, not a lifetime construction cost.

\subsection{Consumer use and source review}
The consumer cohort comprises the first twelve new releases in verification order, with first accepted versions pinned. Separate Fable 5.2 agents under the same operator could modify XGBoost~\cite{xgboost} features, preprocessing, and parameters using training and labeled validation data within the predictor contract. Each release received three 85-minute attempts. Models were frozen before isolated-row held-out scoring; holdout access during research was procedurally prohibited. Provider failures and one premature stop were retried with unchanged pins and budgets; original records remain available.

We define \emph{usable} as supporting a completed prediction experiment under the predictor contract, ending in held-out scoring of a frozen model. This definition does not establish task validity.

For predictive performance, we compare consumers with an untuned XGBoost reference using 100 trees, depth 6, learning rate 0.1, and native categorical handling. A random-search control~\cite{randomsearch} tunes tree and sampling parameters on the same permitted inputs. Each control targets its paired consumer's fitting CPU from successful primary validation trials, subject to a wall-time cap and final-trial overshoot; inference, backtests, and other diagnostics are excluded. Consumers and controls select models by validation area under the ROC curve (AUC); final scoring uses held-out data to separate selection from evaluation~\cite{cawley}. Repeated-run studies motivate reporting variation~\cite{identical-runs}. We plot median paired AUC differences and ranges across three runs. Ranges describe observed run variation, not sampling uncertainty.

Bounded Fable reviews examined pinned task descriptions, construction code, and permitted public data before scoring for target meaning, timing, selection, splits, and label-derived history. Instructions were shared; publisher-documentation access varied. Raw snapshots and holdout rows were outside scope. Table~\ref{tab:releases} distinguishes contradictions, concerns, and conditions of use. Disclosure checks compare findings with pinned READMEs and construction descriptors; \emph{not found} refers to these documents. Explicit isolated-row access counts as a stated restriction. The last two releases were outside both evaluation cohorts.

\begin{figure}[t]
\centering\includegraphics[width=\columnwidth]{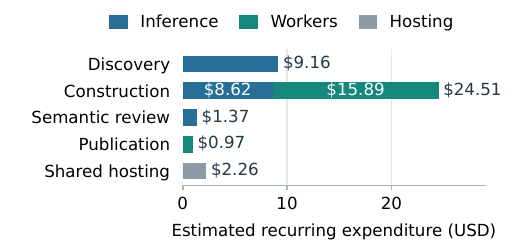}
\caption{Estimated expenditure by stage at midpoint inference tariffs; construction includes checks and baseline fitting.}
\label{fig:costs}
\end{figure}

\begin{figure*}[t]
\centering\includegraphics[width=\textwidth]{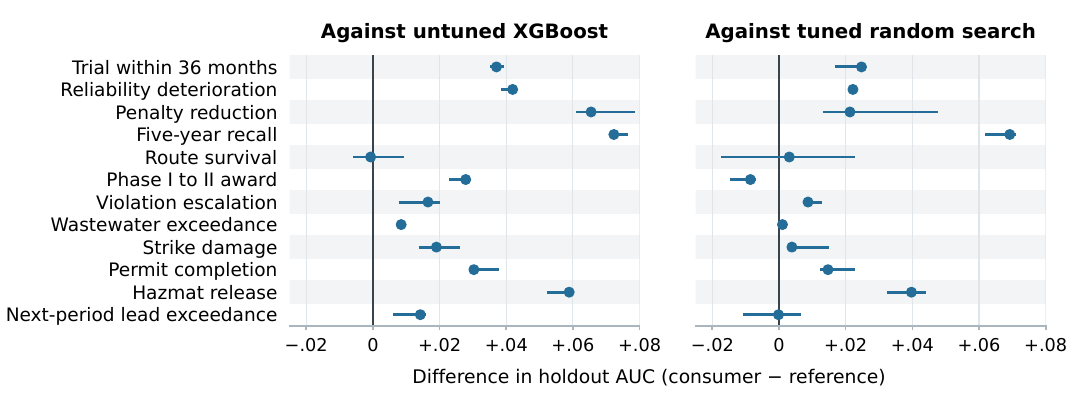}
\caption{Consumer AUC minus each reference. Points show median paired differences; bars span three runs. Positive values favor consumers.}
\label{fig:consumer}
\end{figure*}

\section{Deployment Results}
\subsection{Operation and recurring expenditure}
Fourteen releases were published and anonymously verified on both platforms. Fifty-four candidates completed discovery, eighteen a build, and fourteen acceptance and publication. At cutoff, those 54 candidates comprised 14 published, 37 deferred, two rejected, and one active. These are continuing-operation milestones, not a closed-cohort conversion rate. Among published candidates, admission-to-publication time ranged from 40 to 231 minutes, with a median of 109.

Of the 37 deferrals, 34 followed build failure limits, two semantic repair limits, and one an explicit build deferral. Limit-triggering attempts retain Retry or Repair outcomes in Table~\ref{tab:attempts}.

Table~\ref{tab:attempts} accounts for all 540 attempts. Its 366 retries comprise 206 worker outcomes, 98 model-invocation failures, 61 proposal/controller validation failures, and one publication reconciliation. The worker outcomes span 206 distinct workers: 205 recorded construction/qualification failures and one expiry before upload. The open attempt was waiting on worker reconciliation. Semantic repairs can require another successful build of the same candidate.

\begin{table}[t]
\caption{Stage attempts by outcome at cutoff. A candidate can complete a stage more than once.}
\label{tab:attempts}
\centering\small
\setlength{\tabcolsep}{2.3pt}
\begin{tabular}{@{}lrrrrrrr@{}}
\toprule
Stage & Done & Retry & Repair & Defer & Reject & Open & Total\\\midrule
\csname @@input\endcsname generated/attempt-rows.tex
\bottomrule
\end{tabular}
\end{table}

Construction stopped when reservations consumed the allowance after workers were deleted. Reconciling reservations with completed worker lifetimes restored construction from saved candidate state.

Estimated recurring expenditure was \$38.23--\$38.33: \$19.10--\$19.21 for inference, \$16.86 for workers, and \$2.26 for hosting. The operating ratio was \$2.73--\$2.74 per publication. All-peak inference pricing raises it to \$3.92. Figure~\ref{fig:costs} attributes expenditure to recorded stages.

\subsection{Consumer use and predictive performance}
All twelve evaluated releases supported completed prediction experiments through held-out scoring and received separate bounded source review. That review identified one selection contradiction, concerns, and conditions of use (Table~\ref{tab:releases}); the number of semantically valid releases remains unknown.

Figure~\ref{fig:consumer} characterizes modeling outcomes within the released contracts. All three consumers exceeded the untuned reference on eleven tasks; \emph{Route survival} was mixed. Paired comparisons with random search favored consumers in all three runs on nine tasks, were mixed on \emph{Route survival} and \emph{Next-period lead exceedance}, and favored random search on \emph{Phase I to II award}. The selection and timing findings in Table~\ref{tab:releases} remain material to their use.

Five of the six listed conditions of use were stated in the releases, including isolated-row access for both history-feature tasks. Carrier-code exits were disclosed, but apparent entries were not. Across eight concern entries, including three for \emph{Next-period lead exceedance}, one was stated, three were partly stated, and four were not found in the inspected documents.

\subsection{A population contradiction survived acceptance}\label{sec:validity}
The \emph{Phase I to II award} task uses SBIR records to predict whether a firm's first Phase I award is followed by a Phase II award from the same agency within four fiscal years. Its published selection rule states: ``no outcome field is used to select rows.'' Yet the builder loops over \texttt{range(year + 1, year + 5)} and tests whether each future year occurs in \texttt{unknown\_phase.get((key, agency), set())}. A match sets \texttt{ambiguous = True}; the subsequent \texttt{continue} excludes that firm's row. Thus later award-phase information determines inclusion. Other filters exclude ambiguous phase or agency records. The release's limitations disclose some exclusions, contradicting its selection rule; their effect on scores is unmeasured.

The original reviewer received the complete builder, population description, passing checks, and instructions to compare code with population rules, but not the full published selection declaration. Its response states ``Population matches the rule'' and ``unknown/missing phases in a window are excluded rather than labeled negative.'' It accepted the task with no findings or repair request. A separate pre-scoring review identified the contradiction from the published code and declaration. The original packet and response establish both incomplete review inputs and approval of the exclusions; they don't establish whether supplying the missing declaration would have changed acceptance.

A different condition appears in \emph{Wastewater exceedance}. A prior-year outcome is a legitimate input when available, but future rows can expose earlier labels through that feature. Our scorer supplies one row at a time; no leakage or consumer misconduct was observed in this scoring. Unrestricted batch use was not established safe.

\section{Engineering Lessons}
\textbf{Retain attempt and resource state} to expose repeated work and resume after worker-reservation failures.

\textbf{Budget for execution as well as inference.} Thirty-five of 37 deferred candidates never completed a build. Construction, checks, and baseline fitting consumed 64\% of expenditure (Fig.~\ref{fig:costs}), with worker charges exceeding inference charges.

\textbf{Review the claims that will be published.} For \emph{Phase I to II award}, checking the builder and a partial descriptor did not reconcile the selection declaration with its filters. We propose supplying the complete population/selection contract and requiring each inclusion rule to match it. This response has not been tested as a remedy.

\textbf{Specify the scoring interface.} The history feature in \emph{Wastewater exceedance} shows why accessible rows and outcomes belong in the task contract. Successful execution and legitimate per-row features do not establish safe use under every interface.

\section{Related Work}
AutoData~\cite{autodata-web} and DatasetResearch~\cite{datasetresearch} address web collection and dataset discovery. MLE-Smith~\cite{qiang2026mlesmith} constructs ML tasks from supplied datasets; AutoSDT~\cite{autosdt} searches repositories and adapts scientific programs into tasks. Their reported per-task costs use different output and accounting boundaries, so they cannot support an efficiency ranking with this deployment. Our contribution is operating evidence: source-to-public-release execution, expenditure including failed work, recovery from an operational fault, and a reconstructed acceptance failure.

MLAgentBench evaluates agents performing ML experiments~\cite{mlagentbench}; our consumers characterize outcomes within the factory's released contracts. Studies of deployed agents emphasize reliability~\cite{production}, while AgentOps organizes lifecycle observability~\cite{agentops}. Jacobs and Wallach distinguish intended constructs from operational definitions~\cite{measurement}; the Agentic Benchmark Checklist separates task validity, outcome validity, and reporting~\cite{abc}. Datasheets connects provenance with intended use~\cite{datasheets}, and AutoDataBench examines generated task quality~\cite{autodatabench}.

\section{Limitations and Evidence}
This is one factory, one production-model family, and a short observation window. Public sources may have appeared in model training, and holdout isolation was procedural rather than technical. Source reviews were bounded and do not establish a defect rate across releases. Current-vintage records restrict historical forecasting claims: \emph{Trial within 36 months} inputs can postdate filing, \emph{Strike damage} and \emph{Hazmat release} use retrospective reports, and field-update timing remains unresolved for \emph{Permit completion}. Human policy, development, and responsibility remain outside the automated dataset decisions.

Release pins, operating records, transcripts, reviews, frozen models, and analysis code are retained in the study repository for public release at submission.\footnote{\url{https://github.com/earino/factory-prospective-study}}

\section{Conclusion}
Dataset Factory connected discovery to public release at about \$2.74 per publication in estimated recurring expenditure. Separate agents used twelve releases, while source review exposed a population contradiction missed by acceptance. The experience supports checking task meaning separately from execution and consumer use.

\section*{AI Assistance}
AI systems assisted implementation, literature search, review briefs, source and transcript review, analysis, and drafting. Production and consumer models are identified above. The authors are responsible for the claims and final text.

\bibliographystyle{IEEEtran}
\bibliography{refs}
\end{document}